\documentclass[12pt,times]{article}
 
\normalsize

\usepackage{graphicx} \usepackage[dvips]{color} 

\begin{document} 

\title{Branes, Charge and Intersections}
\author{Donald Marolf}
\maketitle

\centerline{Physics Department, Syracuse University, Syracuse,
         New York 13244}
\begin{abstract}
This is a brief summary of lectures given at the Fourth Mexican School on
Gravitation and Mathematical Physics.  
The lectures gave an introduction to branes in eleven-dimensional
supergravity and in type IIA supergravities in ten-dimensions.  Charge
conservation and the role of the so-called `Chern-Simons terms' were
emphasized. Known exact solutions were discussed and used to 
provide insight into the question 
`Why don't fundamental strings fall off of D-branes,' which is often
asked by relativists.  The following is a brief overview of the lectures
with an associated guide to the literature.
\end{abstract}    

\section{Preface}

This course was intended to be similar to the set of lectures I
gave introducing branes in supergravity and string theory at the 
Third Mexican School on Gravitation and Mathematical Physics in 1998.      
A full write up of this previous set of lectures can be found in
\cite{maz}.  The presentation in \cite{maz} is in fact more similar
to the lectures given at the Fourth School in 2000 than to the original
1998 lectures and I would still recommend \cite{maz}
for an introduction to the subject.  As with the 1998 lecture series, the
style was intended for an audience with a relativity background as opposed
to other introductions aimed more at those with a particle physics background.
I apologize for the fact that several typographic errors remain in
\cite{maz}, though I believe that the equations are now correct.
Readers of \cite{maz} are encouraged to e-mail me at marolf@physics.syr.edu
to point out such errors and make suggestions for future versions that will
eventually be produced.

The more recent lecture series went beyond the material in \cite{maz}
by including discussions of the so-called `Chern-Simons terms.'  These are
terms in the supergravity actions that are responsible for the interesting 
features that arise in more complicated cases such as `intersections of
branes.'  In particular, the goal
was to include some more recent results from \cite{branetrans,frad,T-dual}.
Unfortunately, this required that the basic introduction be shortened in
the 2000 lecture series which likely made these lectures more difficult
for the uninitiated.  The reader interested in learning this additional
material will surely benefit from taking the time to digest \cite{maz}
in full before studying \cite{branetrans,frad,T-dual}. These latter
papers were written as research papers and not as pedagogical introductions
or reviews.  Though I believe that \cite{maz} does contain sufficient
introduction to allow the reader to follow \cite{branetrans,frad,T-dual},
the reader is advised to read these latter papers somewhat more slowly
than \cite{maz}.  Note, by the way, that I do not mean to say
that \cite{maz} will be a quick read for those who are new to the subject.

Below, I give a brief summary of the material covered in the 2000 lecture
series.
I am afraid that this summary is little more than a list of topics covered
together with a somewhat more comprehensive guide to the literature.
Nevertheless, I hope that it will be of use.

\section{Summary}

The 2000 lecture series began with an introduction to eleven-dimensional
supergravity.  Given certain generally accepted caveats, there is
in fact a unique supergravity theory in eleven-dimensions and it was
first obtained in \cite{CJS}.  With the same caveats, eleven is
in fact the highest number of dimensions in which a supergravity theory
exists.  For reasons related to this observation, it turns out the
eleven-dimensional supergravity is the simplest point
for a physicist trained in General Relativity to begin to learn about
string theory and supergravity.  The point is that, to a first 
approximation, the dynamics of
eleven-dimensional supergravity are essentially those of an Einstein-Maxwell
theory (together with some Fermions).  
In particular, eleven-dimensional supergravity contains no
`dilaton' field as do other relevant supergravity theories.  
Theories with a dilaton are not minimally coupled in the sense of the strong
equivalence principle and, as a result, hold a few extra surprises
for relativists.

The idea of the lectures was to start by thinking of supergravity
as being much like Einstein-Maxwell theory and then to slowly add back
the features that distinguish it.  The first such property is that while
the familiar Maxwell field has a rank two field strength tensor $F$, 
the supergravity gauge fields have field strengths  which are
$p$-forms (i.e., rank $p$ covariant anti-symmetric tensors), each with a 
different value of $p$.  It is the feature $p> 2$
which leads to the introduction of
`branes.'  `Brane' is a word for an extended object and is derived from
the word membrane.  In modern terminology, a membrane is known as a `2-brane'
because of it is extended in two spacelike directions; that is, it is 2+1
manifold.  Similarly, strings are known as 1-branes and particles as 0-branes.
Higher dimensional branes also arise in string theory and supergravity.

It turns out that the fundamental electric charges of rank $p$
gauge fields for $p > 2$ are necessarily such extended objects and that
particles are necessarily neutral under such gauge fields.   
The details (as well as a similar discussion 
for magnetic charges) can be found in \cite{maz}.  Indeed, a large part
of \cite{maz} is devoted to this point.  

One may either consider these
gauge fields alone or one may include their couplings to gravity.
In this latter case, one finds an associated set of so-called charged
black $q$-brane solutions.  These solutions are 
analogs of Reissner-Nordstr\"om \cite{nord}
black holes but with horizons that are extended in $q$ directions
instead of being compactly generated.   
For simplicity, these solutions are often discussed only in the extremal
limit and this was the case both in the lectures and in \cite{maz}.  
In eleven-dimensional
supergravity these black branes have smooth horizons even
in the extremal limit just as in Einstein-Maxwell theory, though this
property fails to hold in most other supergravity contexts.
Readers interested in the non-extremal solutions should consult other
reviews of black branes in string theory such as \cite{YR,Stelle,Duff,Peet}.

Another complication of eleven-dimensional supergravity is of course
the Fermions needed for supersymmetry.  While I did not address the 
Fermions in the 2000 lecture series, one may find a brief introduction
to their properties in \cite{maz}.  For more details, the reader
may wish to consult \cite{CJS,GSW,Joe,GH}.

The final complication arises from the so-called Chern-Simons term.
This term has certain features in common with the distinctive term
in 2+1 Chern-Simons theory, which is of course a topological
field theory having no local degrees of freedom.  However, the effects of
this term discussed in the lectures have little to do with topological
quantum field theory.  

To understand just what these effects are, it turns out 
to be convenient to postpone a discussion of the Chern-Simons
term until after discussing how ten-dimensional supergravity arises
from Kaluza-Klein reduction of the eleven-dimensional theory.
Thus, Kaluza-Klein reduction was discussed next in the lectures and is
the next topic in \cite{maz}.  It is important to have some understanding
of Kaluza-Klein reduction before moving on to the new material not
included in \cite{maz}.  In the lectures and \cite{maz} I consider
only Kaluza-Klein reduction on circles, but \cite{MS} is a standard
reference for more general reductions.

The reason that it is easier to first discuss Kaluza-Klein reduction and
to only later address Chern-Simons terms is that
the reduction process in fact creates
additional Chern-Simons terms.  This may seem like an additional
complication, but on turning the picture around it yields
insight into generic Chern-Simons terms.  The point is that
one may find a geometric picture of why the Chern-Simons terms arise
in the dimensional reduction and this geometric picture clarifies
properties of generic Chern-Simons terms.   In particular, the effect
of these `geometric' Chern-Simons terms on intersections of branes
is more easily seen as the effect of the twisting of space in the eleventh
dimension on a simple configuration of branes in eleven dimensions.
This point is not addressed in \cite{maz}, though it forms the main
theme of \cite{frad}.  

A particular example is considered in
detail in \cite{branetrans}, including a lower dimensional example that
is much more easily visualized.  The discussion of \cite{branetrans}
is more elementary and may make a better starting point, though \cite{frad}
addresses additional issues such as charge quantization.  See also
\cite{Town} for an earlier and rather general discussion of Chern-Simons
terms and brane intersections.

The discussion above covers the main general topics 
addressed in the 2000 lecture series.  In addition to describing this
general structure, a final goal of the course was to describe how these
general properties could be used to extract information about non-perturbative
physics in string theory.  The example given in the lectures considered 
fundamental strings attached to a particular type of D-brane, the famous
branes on which fundamental strings are allowed to end.  For an introduction
to these branes the reader should consult the review \cite{clifford} or
the text \cite{Joe}.

The focus of this final discussion concerns 
the sensible question ``what
prevents fundamental strings from falling off of (i.e., separating from)
D-branes and drifting off on their own?''  I have
often heard relativists ask this question of string theorists.
A common answer to this
question invokes charge conservation, though this answer can be confusing
to relativists because the charge considered is not one 
that is seen in supergravity.  Indeed, strictly speaking this common answer
turns out to be true only in the setting of perturbative string theory.

Nevertheless, a related answer can be obtained by studying the conservation
of a supergravity charge.  One finds that the fundamental string
{\it can} in fact separate from the D-brane, but only by transforming
itself into a higher dimensional brane which, at least in the
perturbative string limit, is in fact much more massive.
In other words, there are energetic reasons for the fundamental string
to remain bound to the D-brane and in the perturbative string limit
this binding is very tight indeed.  On the other hand, the binding is
rather weak at large values of the string coupling so that at the
non-perturbative level there can be significant fluctuations away from
the D-brane.
While this picture follows from general reasoning, the solutions
constructed in \cite{branetrans} and the related constructions of
\cite{CGS,CGST} give a closely connected concrete example.
The reader interested in more details should consult
the second section of \cite{T-dual}, in particular in the material associated
with figure 1 of that reference.  The treatment there is somewhat
brief as the point is not central to that paper, but unfortunately
I know of no other discussions of this point.

The reader who succeeds in absorbing the information outlined above
will have gone a long way toward being able to understand discussions
of branes in supergravity and even toward beginning research projects
of their own.  Although I have described it only briefly here, there
is in fact a sizable amount of information to learn.  I should state
that the review \cite{maz} based on my 1998 lectures addresses a number
of additional topics that I was not able to include in the 2000 lecture
series and that I have not mentioned in this summary. The reader pursuing
a broad understanding of branes in supergravity should certainly
digest such additional material as well, though those desiring only
a mild introduction will be sufficiently occupied with the topics listed here.

New reviews of branes and supergravity appear on a regular basis, and I
encourage the reader to scan the archives for further resources.
I expect that a fully revised and expanded version of \cite{maz} will be
available before too many more years pass, but I hope that the present summary
and guide to the literature can be of some use until this does in fact occur.


\end{document}